\documentclass[a4paper,onecolumn,11pt,noarxiv]{quantumarticle}

  \usepackage[utf8]{inputenc}
  \usepackage[english]{babel}
  \usepackage[T1]{fontenc}

  \usepackage{amsmath,amsthm,amsfonts,amssymb,dsfont,graphicx,mathtools,physics,thmtools,thm-restate}
  \usepackage{xprintlen}

  \usepackage{complexity}
  \usepackage{enumitem}
  \usepackage{fullpage}

  \usepackage{comment}
  \usepackage{algorithm}  
  \usepackage{algpseudocode}
  \algrenewcommand\algorithmicrequire{\textbf{Input:}}
  \algrenewcommand\algorithmicensure{\textbf{Output:}}
  \usepackage[most]{tcolorbox}
  \definecolor{pink}{RGB}{237,16,118}
  \definecolor{applegreen}{rgb}{0.55, 0.71, 0.0}
  \colorlet{boxcolor}{applegreen!90!black}
  \newtcolorbox{pinkbox}{%
    colback      = pink!10!white,
    colframe  = pink,
    opacityframe = 1,
    enhanced,
  }

  \newtcolorbox{mybox}[1]{%
  breakable,
    colback      = boxcolor!10!white,
    colbacktitle = boxcolor!50!white,
    colframe  = white,  
    sharp corners,
    opacityframe = .8,
    enhanced,
    title = \color{black} \sffamily #1
  }
  \usepackage{xcolor}
  \usepackage{hyperref}
  \hypersetup{
    colorlinks=true,
    hypertexnames=false,
    linktocpage,
    colorlinks=true, 
    urlcolor=magenta!90!black,    
    linkcolor=blue!60!black, 
    citecolor=black!60 
  }
  \usepackage[capitalize, compress]{cleveref}
  \usepackage[
    backend=biber,
    citestyle=alphabetic,
    bibstyle=alphabetic,
    minalphanames=3,
    maxalphanames=4,
    maxnames=10,
    minnames=1,
    maxcitenames=3,
    mincitenames=1,
    eprint=true,
    date=year,
    doi=true,
    url=false,
    isbn=false,
  ]{biblatex}
  \AtEveryBibitem{%
    \ifentrytype{online}
      {}%
      {%
        \ifentrytype{misc}
          {}%
          {%
            \clearfield{url}%
            \clearfield{doi}%
            \clearfield{urldate}%
            \clearfield{primaryclass}%
          }%
      }%
  }
\newclass{\stoqma}{StoqMA}
\newclass{\classP}{P}
\newclass{\bqp}{BQP}
\newclass{\qcam}{QCAM}
\newclass{\postbqp}{postBQP}
\newclass{\posta}{postA}
\newclass{\postiqp}{postIQP}
\newclass{\classa}{A}
\newclass{\bpp}{BPP}
\newclass{\fbpp}{FBPP}
\newclass{\pp}{PP}
\newclass{\cocp}{coC_=P}
\newclass{\ph}{PH}
\newclass{\np}{NP}
\newclass{\conp}{coNP}
\newclass{\gapp}{GapP}
\newclass{\approxclass}{Apx}
\newclass{\gapclass}{Gap}
\newclass{\sharpP}{\#P}
\newclass{\ma}{MA}
\newclass{\am}{AM}
\newclass{\qma}{QMA}

\newclass{\hog}{HOG}
\newclass{\quath}{QUATH}
\newclass{\bog}{BOG}
\newclass{\xeb}{XEB}
\newclass{\xhog}{XHOG}
\newclass{\xquath}{XQUATH}
\newclass{\maxcut}{MAXCUT}
\newclass{\sat}{SAT}
\newclass{\maxtwosat}{MAX2SAT}
\newclass{\twosat}{2SAT}
\newclass{\threesat}{3SAT}
\newclass{\sharpsat}{\#SAT}
\newclass{\se}{Sign Easing}
\newclass{\classx}{X}

\usepackage{enumitem}
\newlist{io}{description}{1}
\setlist[io]{style=nextline, font=\bfseries, labelsep=0.6em, leftmargin=2.6cm}

\crefname{problem}{Problem}{Problems}
\Crefname{problem}{Problem}{Problems}

\newtheorem*{task}{Task}
\newtheorem*{regime}{Regime}
\newtheorem{scenario}{Scenario}
\newtheorem*{goal}{Goal}

\crefname{theorem}{Theorem}{Theorems}
\Crefname{theorem}{Theorem}{Theorems}
\crefname{conjecture}{Conjecture}{Conjectures}
\Crefname{conjecture}{Conjecture}{Conjectures}
\crefname{lemma}{Lemma}{Lemmas}
\Crefname{lemma}{Lemma}{Lemmas}
\crefname{definition}{Definition}{Definitions}
\Crefname{definition}{Definition}{Definitions}

\theoremstyle{definition}                     
      
\crefname{fact}{Fact}{Facts}
\Crefname{fact}{Fact}{Facts}

\crefname{proposition}{Proposition}{Propositions}
\Crefname{proposition}{Proposition}{Propositions}

\newcommand{\mb}{\mathbb}

\usepackage[most]{tcolorbox}

\newcommand{\id}{\mathbbm{1}}

\DeclareMathOperator*{\Eb}{\mb E}

  \newcommand{\simons}{%
  Simons Institute for the Theory of Computing, University of California at Berkeley}
  \newcommand{\ethz}{%
  Institute for Theoretical Physics, ETH Z\"urich}

\begin{document}

  \title{Has quantum advantage been achieved?}
  \author{Dominik Hangleiter}
  \affiliation{\simons}
  \affiliation{\ethz}
  \date{\today}

  \maketitle
  \begin{abstract}
    Quantum computational advantage was claimed for the first time in 2019 and several experiments since then have reinforced and strengthened the claim. At the same time, a new generation of quantum computing devices with 100 logical qubits is being built. This raises two questions: Has quantum advantage actually been achieved? And what should our next milestones be for the upcoming 100-logical-qubit era? In this perspective, I argue that, in fact, quantum advantage has been achieved. The status today is analogous to Bell-inequality violations in the 1980s where some loopholes remain open, specifically, scalability and verifiability. I then outline three milestones for the 100-logical-qubit era aiming to close those loopholes: demonstrate fault-tolerant quantum advantage, perform efficiently verifiable advantage using random circuits with symmetries, and eventually demonstrate classically verifiable advantage with applications to certified randomness. These milestones are also natural stepping stones towards running algorithms with cryptographic applications. 
  \end{abstract}



  \section*{Introduction}

  More than half a decade has passed since the first experimental claim \cite{arute_quantum_2019} of ``quantum supremacy'' as John Preskill called the idea ``to perform tasks with controlled quantum systems going beyond what
  can be achieved with ordinary digital computers'' \cite[][p.~1]{preskill_quantum_2012} back in 2012. 
  These experiments are now moving farther and farther beyond the reach of classical simulation. 
  At the same time, a genuinely new generation of quantum computers is being built right now. Such devices will be based on quantum error correction and may be capable of performing computations with up to a million operations on 100 logical qubits \cite{tripier_fault-tolerant_2026}.  
  It therefore seems like it is the right time to weigh where the field of quantum advantage could or should move on to next, beyond just performing more of the same types of quantum advantage experiments. 

  As I became interested in this question over the past year, I 
  started off by (anecdotally) polling the community---scientists working on quantum computing in theory and experiments---on whether they believed quantum advantage had been achieved yet at all. 
  Just this one, simple question.  
  As someone who has been working on and following the details underlying the debate about quantum advantage, I was certain that the broader community would be overwhelmingly convinced of the successful demonstration of quantum advantage. 
  If anything, so I thought, in their eyes, quantum advantage would have been achieved \emph{even before} Google's claim, when so-called \emph{quantum simulation} experiments claimed to have performed computations that no classical computer could do (e.g.~\cite{choi_exploring_2016}).

  I could not have been more wrong: 
  Less than half of the community seems to believe that quantum advantage has actually been achieved.

  In the discussions that ensued, I came to understand popular criticisms of the experiments that have been performed and even of the concept of quantum advantage to begin with. 
  Most of all, it seemed to me, the community had dismissed Google's advantage claim because of the classical simulation shortly after~\cite{pan_solving_2022,kalachev_classical_2021}. 
  People did not seem to be aware that it was only the first in an impressive series of similar demonstrations that became bigger and better with every year that passed. 
  They also had not kept track of all the theoretical advances in our understanding of quantum advantage since then.

  This is why my first goal in this perspective is to take stock and recap the quantum advantage experiments that have been performed so far. 
  I outline the somewhat subtle argument that has emerged over the past decade why these experiments in fact demonstrate a quantum advantage in spite of the fact that they are noisy. 
  My goal in the first part of this perspective is then to convince you that, in fact, quantum advantage has been achieved. 
  To do this, I narrow down the focus on what I think are the strongest examples of quantum advantage demonstrations.
  These are ``CS-style'' demonstrations, where a clear computational task is being solved, there is strong, complexity-theoretic evidence for a quantum advantage, and simulations using the best algorithms are clearly out of reach for existing classical computers. 
  Doing so allows me to go into the depth necessary to get across why and in what sense these experiments pass the bar for ``quantum advantage''. 
  Let me caution, though, that the task these experiments solve, \emph{random circuit sampling}, is a (nearly) useless one that is based on running entirely unstructured quantum computations. 
  None of the experiments performed so far are relevant to scientifically, commercially, or societally relevant problems beyond the demonstration of quantum advantage itself.

  In particular, I discuss in depth what the precise task is that the experiments demonstrate an advantage on---an issue that turns out to be quite subtle. 
  I then discuss how the experimental samples are validated using the so-called cross-entropy benchmark and how developing an understanding of its correspondence (or not) to the quantum fidelity has been crucial to establishing a solid foundation of the advantage claims.
  I argue that the experiments meet the same evidential standard as most physics experiments and, specifically, as loophole-limited Bell tests, and possibly even a higher standard. 
  I think of the situation we are in now in quantum advantage as being analogous to that of Bell-inequality violations in the late 1980s: 
  we can be confident that quantum computation with an advantage is possible but there are loopholes that remain open for nature itself or just the provider of a quantum computer to cheat if they act adversarially.

  The detailed argument for why and how quantum advantage has been achieved then serves as a stage for my perspective on the challenges in theory and experiment of quantum advantage for the next generation of quantum computers, in the 100-logical-qubit era.
  These are devices with 100 logical qubits on which we can perform up to a million operations \cite[e.g.][]{tripier_fault-tolerant_2026}. 
  My argument will expose the loopholes of the demonstrations, in particular, scalability and verifiability. 
  The goal of the agenda I propose is to close those loopholes step by step.
  Its overall mindset is to move from the completely unstructured circuits to computations with structure, which ideally find a useful application. 

  First, we should demonstrate scalable quantum advantage using logical qubits. This is achievable in the near term using so-called IQP circuits \cite{hangleiter_fault-tolerant_2025}. 
  Second, we should work towards closing the verification loopholes by identifying and performing structured computations whose symmetries can be used to verify the solution of the computational task under reasonable assumptions on the experiment, ideally in a fault-tolerant setting. 
  In fact, first advantage-scale experiments in this direction have been performed very recently \cite{martiel_sampling_2026}.
  Finally, we should aim to close the verification loophole in adversarial settings, where we only have remote access to a quantum computer. 
  There is a big open space for theory research into tailored cryptography here. This might also lead to first, low-cost applications such as certifiable random number generation \cite{aaronson_certified_2023}.

  I argue that while theory challenges remain open, achieving those milestones is not only feasible in the coming years but they also serve as clear stepping stones towards running more structured algorithms and in particular cryptographic applications of quantum computers.
  Further along this path, yet another step beyond the 100-logical-qubit regime, performing Shor's algorithm on cryptographically relevant problems is racing into the reach of feasibility of future quantum computers with low thousands of logical qubits capable of running billions of logical gates~\cite{babbush_securing_2026,cain_shors_2026}. 

  The agenda I propose is a concrete one that is focused on next-generation ``CS-style'' experiments on quantum computers, starting from the quantum advantage demonstrations of the past years and aiming towards future cryptographic applications of quantum computers. 
  In this perspective, I will therefore take you on a deep dive into the field of random circuit sampling and sketch a path for its future. 
  Complementary perspectives, encompassing the breadth of potential quantum advantages have recently been given in Refs.~\cite{huang_vast_2026} and \cite[][Sec.~1]{menssen_strategic_2026}.

  In \cref{sec:part1}, I recap what quantum advantage is and how it has been demonstrated in a set of experiments over the past few years. 
  \cref{sec:part2} then contains the deep-dive into the theoretical underpinning of these experiments. It gives the core argument why quantum advantage has been achieved, and responds to common counter-arguments. 
  Having settled the status quo, in \cref{sec:part3}, I am then ready to give my perspective on what should be done with the next-generation, 100-logical-qubit devices in the field of quantum advantage. 

  \section{What is quantum advantage and what has been done?}
  \label{sec:part1}

  The evidence for realistic quantum advantage dates back to the mid-1990s. 
  Back then, not only did Lloyd and Shor discover two basic quantum algorithms---simulating quantum systems \cite{lloyd_universal_1996} and factoring large numbers \cite{shor_algorithms_1994}---for which there are presumably exponential quantum speedups. 
  It was also shown that quantum computing is a reasonable model of computation \cite{bernstein_quantum_1997} in which errors can be corrected \cite{shor_fault-tolerant_1996} and this can even outweigh the additional errors incurred during this correction process \cite{aharonov_fault-tolerant_1997}. 
  These results show that large-scale quantum computers under plausible error mechanisms would be able to solve problems efficiently that no classical computer could solve. 
  So why talk about quantum advantage 20 and 30 years later? 

  The idea of a quantum advantage demonstration---be it on a completely useless task even---emerged as a milestone for the field in the 2010s. Achieving quantum advantage would finally demonstrate that quantum computing was not just a random idea of a few academics who took quantum mechanics too seriously. It would show that quantum speedups are real. It would show that we can actually build quantum devices, control their states and the noise in them, and use them to solve tasks which not even the largest classical supercomputers could do.

  \subparagraph{What is quantum advantage?}

  But what exactly do I mean by ``quantum advantage''? It is a vague concept, for sure, but the essential criteria of quantum advantage for quantum computation are the following.

  \begin{enumerate}
  \item The quantum device needs to solve a pre-specified computational task. This means that there needs to be an input to the quantum computer. Given the input, the quantum computer must then be \emph{programmed} to solve the task for the given input. This may sound trivial. But it is crucial because it delineates programmable computing devices from experiments on an arbitrary physical system.

  \item There must be a scaling difference in the time it takes for a quantum computer to solve the task and the time it takes for a classical computer. As we make the problem or input size larger, the ratio of classical runtime to quantum runtime should grow, ideally exponentially. 

  \item And finally: the actual task solved by the quantum computer should not be solvable by any classical machine (at the time). 
  \end{enumerate}

  Achieving this last criterion using imperfect, noisy quantum devices is the challenge the idea of quantum supremacy set for the field, theorists and experimentalists alike. After all, running any of our favourite quantum algorithms in a classically hard regime on these devices is completely out of the question. They are too small and too noisy. So the field had to come up with the conceivably smallest and most noise-robust quantum algorithm that has a significant scaling advantage against classical computation. 
  And it had to improve the devices to a level sufficient for quantum advantage. 

  \subparagraph{Random circuit sampling.}

  The idea is simple: we just run a random computation, constructed in a way that is as favorable as we can make it to the quantum device while being as hard as possible classically. This may strike you as quite an unfair way to come up with a computational task---it is just built to be hard for classical computers without any other purpose. But it is a fine computational task. There is an input: the description of the quantum circuit. The device needs to be programmed to run this exact circuit. And there is a task: just return whatever this quantum computation would return. These are strings of 0s and 1s drawn from a certain distribution. Getting the distribution of the strings right for a given input circuit is the computational task. 

  This task, dubbed \emph{random circuit sampling} (RCS), can be solved on a classical as well as a quantum computer, but there is a (presumably) exponential advantage for the quantum computer. More on that in \cref{sec:part2}.

  Let me first tell you about the experimental demonstrations of random circuit sampling. The task solved in random circuit sampling is to produce bit strings $x \in \{0,1\}^n$ distributed according to the Born-rule outcome distribution
  $$p_C(x) = \left| \bra x C \ket {0}\right|^2$$
  of a sequence of elementary quantum operations (unitary rotations of one or two qubits at a time) $C$ which is drawn randomly according to certain rules. This circuit $C$ is applied to a reference state $\ket 0$ on the quantum computer and then measured, giving the string $x$ as an outcome. 

  \subparagraph{The breakthrough: classically hard programmable quantum computations in the real world.}
  In the first quantum supremacy experiment \cite{arute_quantum_2019} by the Google team, the quantum computer was built from 53 superconducting qubits arranged in a 2D grid. The operations were randomly chosen simple one-qubit gates ($\sqrt X$, $\sqrt Y$, $\sqrt{X+Y}$) and deterministic two-qubit gates called $\text{fSim}$ applied in the 2D pattern, and repeated 20 times (the \emph{depth} of the circuit). The limiting factor in these experiments was the quality of the two-qubit gates and the measurements, with error probabilities around 0.6\% and 4\%, respectively. 
  A very similar experiment was performed by the USTC team on 56 qubits also with depth 20 \cite{wu_strong_2021} and both experiments were repeated with lower error rates (0.4\% and 1\% for two-qubit gates and measurements), slightly larger system sizes (70 and 83 qubits, respectively), and depth 32 a few years later \cite{morvan_phase_2024,gao_establishing_2025}.
  In all of these experiments, the devices had limited programmability in that the exact two-qubit gates they performed depended on the specific processor chip and location thereon because of limitations to the calibration of the chips. 
  The gate set therefore did not just depend on the architecture, but the specifics of the physical device itself (see, for example, Fig.~6 in the supplement of \cite{morvan_phase_2024}). 

  Using a trapped-ion architecture, the Quantinuum team also demonstrated random circuit sampling on 56 qubits and depth 20 but with arbitrary connectivity (random regular graphs) \cite{decross_computational_2025}. There, the two-qubit gates were $\pi/2$-rotations around $Z \otimes Z$, the single-qubit gates were uniformly random and the error rates much better (0.15\% for both two-qubit gate and measurement errors). Recently, this experiment has been reproduced on 98 qubits at depth 26 and with even better error rates~\cite{ransford_98-qubit_2026}.
  Importantly, the trapped-ion implementations do not face the gate-calibration issues of the superconducting devices that lead to device-dependent gate sets. 

  All the experiments ran random circuits on varying system sizes and circuit depths, and collected thousands to millions of samples from a few random circuits at a given size. To benchmark the quality of the experiments, they used the now widely accepted \emph{linear cross-entropy (XEB) benchmark} defined as 
  \[
    \chi = 2^{n}\, \mathbb E_C \mathbb E_{x}\, p_C(x) -1 ,
  \]
  for an $n$-qubit circuit. The expectation over $C$ is over the random choice of circuit and the expectation over $x$ is over the experimental distribution of the bit strings.  In other words, to compute the XEB given a list of samples, you `just' need to compute the ideal probability of obtaining each sample from the circuit $C$ and take the mean of those probabilities. 

  The XEB is a good benchmark for random circuits because it gives~$1$ for ideal samples and~$0$ for uniformly random samples, and it can be estimated accurately from polynomially many samples. 
  It thus distinguishes samples obtained from an ideal experiment from trivial uniform samples.
  Under the right conditions, it turns out to be a good proxy for the many-body \emph{fidelity} of the quantum state prepared just before the measurement. 

\begin{figure}[h]
  \centering
  \includegraphics[width=.8\linewidth]{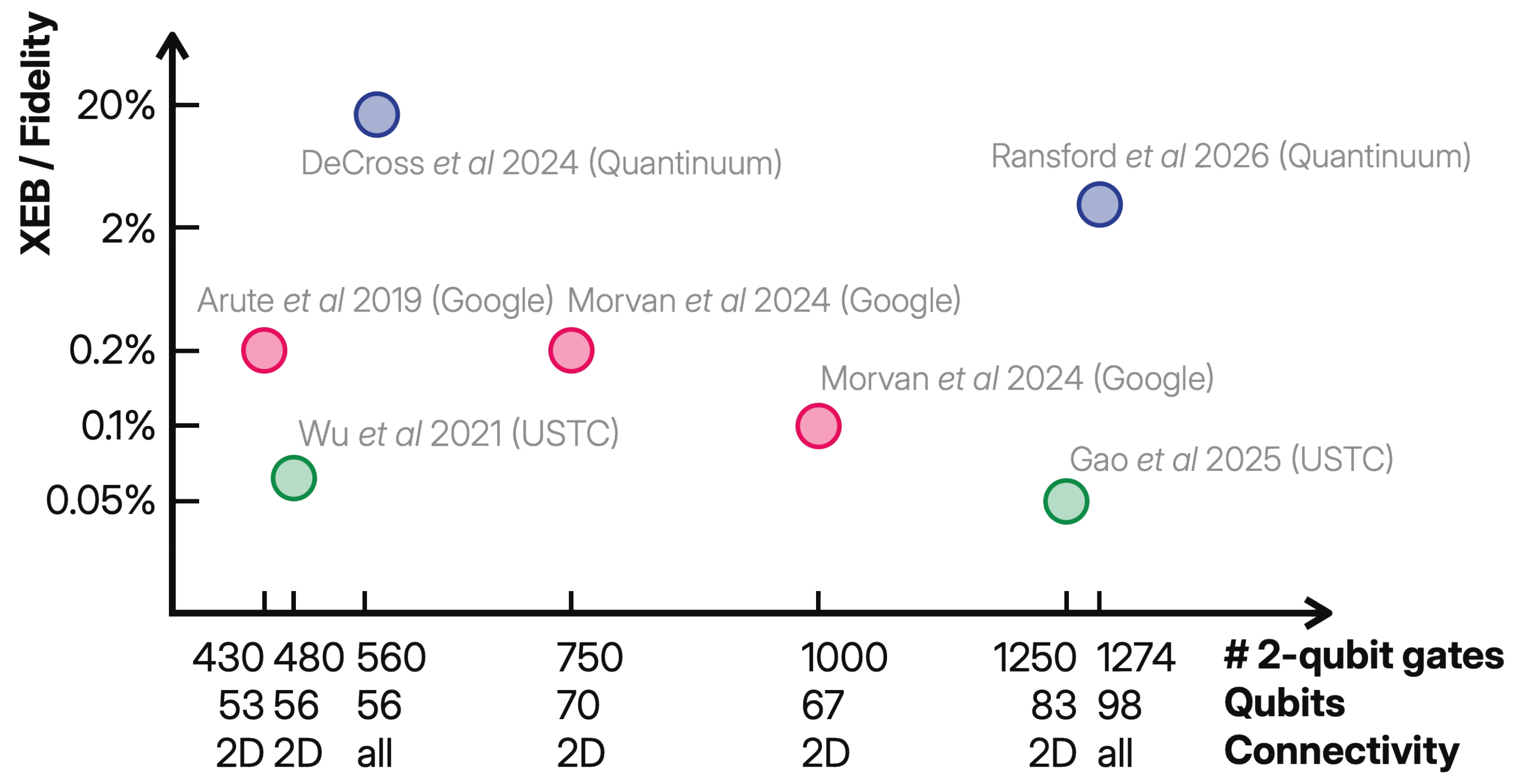}
  \caption{\label{fig:xeb fidelities experiments} XEB/fidelity achieved by the random circuit sampling experiments since 2019.}
  \end{figure}

  We should therefore expect an XEB score of $(1-\text{error per gate})^{\text{\# gates}} \sim e^{- \varepsilon n d }$ for some noise- and architecture-dependent  constant $\varepsilon$. All of the experiments achieved a value of the XEB that was significantly (in the statistical sense) far away from $0$ as shown in \cref{fig:xeb fidelities experiments}. 
  This shows that the experiments achieved a nontrivial value of the XEB compared to a uniform sampler with XEB $0$ or a sampler from an uncorrelated state which would have fidelity $2^{-n}$ with the ideal state. 
  This number is less than $10^{-14}\ $\% for all the experiments. 
  But maybe there are better algorithms than those trivial ones. 

  \paragraph{Classical simulation of the quantum advantage experiments.}
  An ideal RCS experiment is most efficiently simulated using a technique called tensor-network contraction \cite{markov_simulating_2008,gray_hyper-optimized_2021}.
  The complexity of this simulator is governed by the treewidth of the spacetime circuit diagram \cite{markov_simulating_2008} or its minimum bisecting cut \cite{decross_computational_2025}. 
  Typically, this is either the number of qubits (at high circuit depths) or an upper bound on the bipartite entanglement generated by the circuit (at low depths), see Figure 5 of the paper by the Quantinuum team \cite{decross_computational_2025}.
  For the 2019 experiment the best estimates in actual cost are on the order of $10^{18}$ to $10^{19}$ FLOPs for computing an individual outcome probability \cite{gray_hyper-optimized_2021,morvan_phase_2024,gao_establishing_2025}.

  To exploit a target XEB value less than $1$ in such a simulation one can judiciously change the circuit by removing a small number of gates from it in order to make the circuit easier to simulate while retaining a nonzero XEB score compared to the original circuit's output distribution~\cite{pan_solving_2022}. 
  In the tensor-network language, this removes certain contraction paths from the overall sum, which can be done in more general ways as well \cite{kalachev_classical_2021}.
  However, for a target fidelity of $\delta$ this approach only reduces the simulation cost by a factor of $\delta$ \cite{kalachev_classical_2021} and reproducing the low scores between 0.05\% and 0.2\% of the experiments remains extremely hard on classical computers. Producing samples that match the experimental XEB score has only been achieved for the first experiment from 2019~\cite{pan_solving_2022}. 
  This simulation already exploited the relatively low XEB score to simplify the computation. 
  Running these techniques for the slightly larger 56-qubit USTC experiment was estimated to require roughly $1.4 \cdot 10^{22}$ FLOPs (the equivalent of a few days on a machine with thousands of GPUs) \cite{kalachev_classical_2021}.
  In the meantime, significantly more effort has been spent on trying to simulate these experiments \cite[see e.g.][]{liu_validating_2024,zhao_leapfrogging_2025,pan_efficient_2024} but the only one of the experiments which has actually been simulated is the 2019 experiment by the Google team, while the 2021 USTC experiment is in the reach of simulability \cite{liu_validating_2024}. 
  The significant work that has gone into the problem of simulating these experiments has so far failed to go beyond these improvements. 
  I would consider any progress that goes beyond the exponential-in-treewidth scaling of the tensor-network contraction \cite{markov_simulating_2008,gray_hyper-optimized_2021,decross_computational_2025} or the linear scaling in the fidelity \cite{kalachev_classical_2021,pan_solving_2022} to be major progress in simulating random quantum circuits. 

  \subparagraph{Proxy of a proxy of a benchmark}

  Given the hardness of computing outcome probabilities, how do you compute the XEB or fidelity of a random circuit sampling experiment in the advantage regime in the first place? 
  After all, the quantum advantage of RCS is  based on that very hardness. 
  To overcome this problem and get an estimate of the XEB in the advantage regime, the experiments needed to use proxies and extrapolation from classically tractable regimes. 
  This will be important for \cref{sec:part2} of this perspective, where I discuss the evidence we have for quantum advantage, so let me give you some more detail.

  To extrapolate to the fidelity of the largest experimental circuits, one can run smaller circuits of increasing sizes and extrapolate to the size in the advantage regime. Alternatively, one can run circuits with the same number of gates but with added structure that makes them classically simulatable and extrapolate to the size of the advantage circuits. Extrapolation is based on samples from experiments that are different from the quantum advantage experiments themselves in that they are smaller or use differently structured circuits. All of the experiments did this. 
  A separate estimate of the XEB score is based on proxies. An XEB proxy uses the samples from the advantage experiments themselves, but computes a different quantity than the XEB that can actually be computed and for which one can collect independent numerical and theoretical evidence that it matches the XEB in the relevant regime. For example, the Google experiments \cite{arute_quantum_2019,morvan_phase_2024} averaged outcome probabilities of modified circuits that were related to the true circuits but easier to simulate. 
  The Quantinuum experiments \cite{decross_computational_2025,ransford_98-qubit_2026} in addition to some XEB estimates at intermediate circuit sizes did something entirely different. 
  They estimated the fidelity of an advantage experiment by running a different experiment in which they ran the advantage circuit forward and backward, effectively  inverting the circuit on the quantum computer. 
  They could then measure the probability of returning to the initial state, which gives a proxy for the fidelity of an experiment in which you only run the circuit forward. 
  All of the methods used to estimate the XEB of the quantum advantage experiments required some independent verification based on numerics on smaller sizes and induction to larger sizes, as well as theoretical arguments. 

  In the end, the advantage claims are thus based on a proxy of a proxy of the quantum fidelity. This is not to say that the advantage claims do not hold. In fact, I argue in  \cref{sec:part2} that this is just the way science works. I will also tell you more about the evidence that the experiments I described here actually demonstrate quantum advantage and discuss some skeptical arguments.

  \paragraph{Why focus on RCS?} 
  Before moving on to the argument for why these experiments show quantum advantage, let me  note that in describing the quantum supremacy experiments, I focused on random circuit sampling which is run on programmable digital quantum computers. What I neglected to talk about is boson sampling and Gaussian boson sampling, which are run on photonic devices and have also been experimentally demonstrated~\cite{spring_boson_2013,wang_boson_2019,zhong_quantum_2020,madsen_quantum_2022}. The reason for this is that I think random circuit sampling experiments make the most compelling case for quantum advantage: we have seen several separate experiments of basically the same sampling task with slightly different sizes and circuit connectivities (something we cannot say for boson sampling), with the XEB we have a single, clear benchmark (something we cannot say for boson sampling), and the experiments (with the exception of the 2019 experiment) have not been simulated, while all boson sampling experiments except for this very recent one~\cite{liu_robust_2025} have been~\cite{oh_classical_2024}.

  \section{Considering the evidence for quantum advantage}
  \label{sec:part2}

  In the previous section, I discussed the idea of random circuit sampling (RCS) and the experimental implementations thereof. In this section, I discuss the arguments and evidence for \emph{why} I am convinced that the experiments demonstrate a quantum advantage. 

  Recall from \cref{sec:part1} that to assess an experimental quantum advantage claim we need to check three criteria: 
  \begin{enumerate}
  \item Does the experiment correctly solve a computational task?
  \item Does it achieve a scalable advantage over classical computation? 
  \item Does it achieve an in-practice advantage over the best classical attempt at solving the task? 
  \end{enumerate}

  \paragraph{What is the issue?}

  When assessing these criteria for the RCS experiments there is an important problem: The early quantum computers we ran them on were very far from being reliable and the computation was significantly corrupted by noise. How should we interpret this noisy data? Or more concisely: 

  \begin{enumerate}
  \item Is random circuit sampling \emph{still} classically hard even when we allow for whatever amount of noise the actual experiments had? 
  \item Can we be convinced from the experimental data that this task has actually been solved? 
  \end{enumerate}

  I want to convince you that we have developed a very good understanding of these questions that gives a solid underpinning to the advantage claim. Developing that understanding required a mix of methodologies from different areas of science, including theoretical computer science, algorithm design, and physics and has been an exciting journey over the past years. 

  \subsection{The noisy sampling task}

  Let us start by answering the base question. What computational task did the noisy experiments actually solve? 
  Even after all the experiments and theoretical developments, this remains maybe the most tricky question in assessing quantum advantage. 

  Recall that, in the ideal RCS scenario, we are given a random circuit $C$ from some ensemble on $n$ qubits and the task is to sample from the output distribution of the state $\ket C$ obtained from applying the circuit $C$ to a simple reference state. The output probability distribution of this state when we measure every qubit is determined by the Born rule.

  What does a noisy quantum computer do when we execute all the gates on it and apply them to its state? 
  It prepares a noisy version $\rho_C$ of the intended state $\ket C$ and once we measure the qubits we obtain samples from the output distribution of that noisy state in the basis of the measurement. 
  In the experiments, this is just the standard basis, but, importantly, it could be any Pauli basis. This is because a random product unitary can be merged with the last layer of single-qubit gates, which  is random in all experiments we considered. 

  When specifying the task solved by the experiments, 
  we should do this in a way that does not presuppose contingent specifics of the computer we want to show an advantage of. 
  These could be specifics of the state prepared by the quantum computer or the noise that determined it, which we can only learn about once we have access to the particular computer being used. 

  Ideally, we would like to define a task that is based purely on the classical output distribution in a fixed measurement  basis or some simple property of this distribution.
  The most natural way to do this is just to ask that the samples are distributed according to the correct distribution up to some fixed statistical (total-variation) distance. 
  \begin{task}[Standard RCS]
    Given a random circuit $C$, sample from any distribution whose statistical distance to the output distribution $p_C$ of $\ket C$ is at most $\nu$ with high probability over the choice of random circuit. 
  \end{task}
  The problem with this task is that testing whether samples from a given distribution are close to some target distribution requires exponentially many samples \cite{hangleiter_sample_2019}. 
  To experimentally verify TVD, we would therefore need to run the quantum computer exponentially many times. 
  In contrast, the  experiments report the XEB, which can be estimated from polynomially many samples. 
  In fact, all known routes to get a handle on TVD run through the XEB, but require assumptions on the device. 
  A more natural task is therefore to directly use the XEB as a benchmark.
  \begin{task}[X-RCS]
    Given a random circuit $C$, sample from any distribution $q_C$ such that the circuit-averaged XEB with the output distribution $p_C$ of $\ket C$ is at least $\delta$, i.e., $\chi \ge \delta$. 
  \end{task}
  If we just want to consider the classical output distribution in a fixed measurement basis, then this is the most literal interpretation of the task the experiments solved, since the quantity they report is the XEB. 
  However, the evidence we have for the hardness of this task is subtle and very sensitively depends on the choice of $\delta$. 
  We will come back to this later. 

  Finally, in my opinion, the best way to interpret the experiments is based on the observation that in every one of the experiments \emph{there is} some noisy state that can be measured in a chosen basis. 
  We can then define a computational task by fixing how accurate that noisy state preparation is. 
  The natural way to do this is to use the \emph{fidelity}
  $$F(C) = \bra C \rho_C \ket C,$$
  which is just the overlap between the ideal state and the noisy state. The fidelity is 1 if the noisy state is equal to the ideal state, and 0 if it is perfectly orthogonal to it.
  \begin{task}[Finite-fidelity RCS]
  Given a random circuit $C$ and a specified Pauli basis, sample from the output distribution in this basis of a quantum state $\rho_C$ (that only depends on $C$) whose circuit-averaged fidelity with the ideal output state $\ket C$ is at least $\delta$, i.e., 
  \begin{align}
  F \coloneqq \Eb_{C} \bra C \rho_C \ket C \ge \delta
  \end{align}
  \end{task}

  It is worthwhile dwelling on some fine points of this task definition for a bit. 
  The finite-fidelity RCS task insists that the classical or quantum algorithm sample from the output distribution of a \emph{fixed} quantum state $\rho_C$ that only depends on $C$ in any specified Pauli basis. 
  In other words, it requires the sampler to ``commit'' to a state $\rho_C$ that accounts for the output distribution of the measurement in any basis.
  This commitment is crucial to the advantage claim for the following reason: 
  If we allow the sampled state in the finite-fidelity RCS task to depend on the measurement basis, or equivalently, if we just ask for the standard-basis distribution of a quantum state with fidelity $\delta$, then trivial uniform sampling achieves the task for any $\delta \lesssim \pi/4$.  
  This is because for the output state of a random circuit there is a closeby phase state which typically has fidelity $\pi/4$ and whose  standard-basis output distribution is just uniform.\footnote{If we write $\ket C = \sum_x \sqrt{p_C(x)} e^{i \varphi_C(x)} \ket x$, there is a closeby phase state $\ket{\phi(C)} = 2^{-n/2} \sum_x e^{i \varphi_C(x)} \ket x$ which can be shown to have fidelity approximately $\pi/4$ with $\ket C$.}

  \paragraph{Classical competition and our threat model of nature}
  The precise definition of the task is important because depending on which task we talk about, the competing algorithms are constrained in different ways. 
  In turn, this will affect whether we can justifiably claim quantum advantage. 
  The different definitions also assume different ``threat'' models for the quantum device. 
  To interpret the experiments for the purely sample-based definitions, we can simply treat the computing device as a black box.   
  For finite-fidelity RCS we must assume that the experiments actually produced a quantum state that does not depend on and can be probed in any (subsequent) choice of measurement basis. 
  In other words, we must assume that nature (or the experimentalist) is not actively trying to fool us. 

  As physicists, we make this assumption every day, and it is the foundation of our ability to formulate laws of nature. 
  We try and understand nature by probing its behaviour in experiments, and a cornerstone of physics is that these experiments are reproducible in different places, on different devices and run by different people. 
  As a cryptographer, you might be more wary---what if someone was trying to sell me a non-functioning quantum computer for one that can achieve quantum advantage. Under this threat model you will be less happy to assume that we are just probing nature in the experiments. \\ 

  I will first give the argument for why the experiments convincingly demonstrate quantum advantage for finite-fidelity RCS.
  For this argument, consider yourself a physicist. 
  Having done so, we will come back to X-RCS and I will give you an argument why the experiments may have even demonstrated advantage for this task. 
  If you believe this second argument, even as a cryptographer, you will have a reason to be convinced of noisy quantum advantage.

  \paragraph{The experiments and finite-fidelity RCS}
  At the most basic level, the noisy experiments' goal was to prepare a really complex, hard-to-simulate quantum state with finite fidelity.
  They reported average XEB or fidelity values of $\delta$ around 0.1\% on superconducting qubits and around $10\%$ on ion traps. What is more, they consistently achieve this value even as the circuit sizes are increased in the later experiments as we saw in \cref{fig:xeb fidelities experiments} (and see also \cref{fig:xeb fidelity comparison} below). Both the actual value and the scaling will be important later when we assess whether the data supports a successful achievement of finite-fidelity RCS.

  The only wrinkle in thinking of the experiments' claims in terms of finite-fidelity RCS  is that they only produced samples in one basis for every random circuit. 
  So did they actually ``commit'' to a state? 
  For this, we need to take it for granted that they \emph{could have} sampled the same state in a different basis with the same sample quality.
  As physicists, we should be happy to do this because of two reasons. 
  First, the experiments extensively characterized the individual gates, and single-qubit gate error rates (single qubit gates are required for the change of measurement basis) were very small compared to the error rates of the entangling gates.
  Second, in the entire noise characterization of the experiment, there was no indication of any global noise that could make the state preparation quality dependent on gates applied \emph{later in the circuit}!
  Taken together, these observations are evidence that the prepared  state could just as well have been measured in any other Pauli basis without  loss in fidelity.   
  Let me also note that a very recent experiment \cite{liu_certified_2025-1} can be literally interpreted as solving finite-fidelity RCS: In those experiments the measurement basis was chosen \emph{only after} the state was prepared, so the quantum computer indeed committed to the state. 

  Note that the notions of RCS we consider (X-RCS and finite-fidelity RCS) do not demand success for \emph{every} circuit, but only \emph{on average} over the random circuit ensemble. 
  This matches what the experiments do: they draw random circuits and need the device to perform well on average over those draws. Accordingly, when the experiments quote a single number as XEB or ``fidelity'', it is really the circuit-averaged fidelity $F$ or XEB $\chi$.

  \paragraph{Quantum advantage of finite-fidelity RCS.}

  What is the complexity of finite-fidelity RCS? 

  Let us start off by supposing that the quantum computation is (nearly) perfectly executed, so the required fidelity $\delta$ is quite large, say, 90\%. In this scenario, we have very good evidence based on computational complexity theory that there is a scalable and in-practice quantum advantage for RCS.
  This evidence is very strong, comparable to the evidence we have for the hardness of factoring and simulating quantum systems. The intuition behind it is that quantum output probabilities are extremely hard to compute because of a mechanism behind quantum advantages: destructive interference. 
  The requirement that the underlying state does not depend on the measurement basis only makes the task harder.  
  This hardness very plausibly even persists for X-RCS, i.e., if we only ask to produce samples that achieve a near-ideal value of the XEB \cite{aaronson_complexity-theoretic_2017}. 
  If you are interested in the subtleties and the open questions, see~\cite{hangleiter_computational_2023}.

  How far down in fidelity does this classical hardness persist? Intuitively, the smaller we make $\delta$, the easier finite-fidelity RCS should become for a classical algorithm (and a quantum computer, too), since the freedom we have in deviating from the ideal state in our simulation becomes larger and larger. This increases the possibility of finding a state that turns out to be easy to simulate within the fidelity constraint. 
  Somewhat surprisingly, though, the classical simulation cost only reduces by a factor of $\delta$ in the best available  simulations of the first Google experiment \cite{kalachev_classical_2021,pan_solving_2022}. 
  Importantly, the tensor-network algorithms used in those simulations in fact solve finite-fidelity RCS even though they only reported outcomes in a single basis just like the experiment. 
  This is because the bulk of the underlying tensor network represents a basis-independent state $\rho_C$ and the choice of measurement basis only affects a few tensors at the boundary of the tensor network. This means that the basis choice does not significantly impact the contraction complexity. 
  All but those two experiments are far out of reach for these algorithms, however.

  There is therefore no known efficient classical algorithm that achieves the finite-fidelity task for $\delta$ inverse polynomially away from the baseline trivial value of $2^{-n}$. This is the value a maximally mixed or randomly picked state achieves because a random state has no correlation with the ideal state (or any other state), and $2^{-n}$ is exactly what you expect in that case (while 0 would correspond to exact orthogonality). 

 

  \paragraph{Getting the scaling right: weak noise and low depth}

  So what is the right value of $\delta$ at which we can hope for a scalable and in-practice advantage of RCS experiments? 

  When thinking about this question, it is helpful to keep a model of the circuit in mind that a noisy experiment runs. Consider a noisy circuit on $n$ qubits with $d$ layers of gates and single-qubit noise of strength $\varepsilon$ on every qubit in every layer. In this scenario, the average fidelity with the ideal state will decay as $F \sim (1-\varepsilon)^{nd} \sim \exp(- \varepsilon n d)$. 

  Any reasonably testable value of the fidelity needs to scale as $1/\mathsf{poly}(n)$, since eventually we need to estimate the average fidelity $F$ from the experimental samples and this typically requires at least $1/F^2$ samples, so exponentially small fidelities are experimentally invisible. Inverse polynomial fidelity $\delta$ is also much closer to the near-ideal scenario ($\delta \geq$ 90\%) than the trivial scenario ($\delta = 2^{-n}$) in terms of complexity. While we cannot formally pin this down, the intuition behind the complexity-theoretic evidence for the hardness of near-ideal RCS persists into the $\delta \sim 1/\mathsf{poly}(n)$ regime: to sample up to such high precision, we still need an inverse-exponentially accurate estimate of the ideal probabilities, and getting such an estimate is computationally extremely difficult. More precisely, it is \sharpP-hard. Scalable quantum advantage in this regime is therefore quite a safe bet. 

  How do the parameters of the experiment and the RCS instances need to scale with the number of qubits $n$ to experimentally achieve the fidelity regime? The limit to consider is one in which the noise rate decreases with the number of qubits, while the circuit depth is only allowed to increase very slowly. It depends on the circuit architecture, i.e., the choice of circuit connectivity, and the gate set, through a constant $c^*$ as I explain in more detail below. 

  \begin{regime}[Weak noise and low depth scaling] \quad  
  \begin{tabbing}
  \emph{Weak noise:} \, \= The local noise rate of the quantum device scales as $\varepsilon < c^*/n$.\\
  \emph{Low depth:} \> The circuit depth scales as $d \lesssim \log n$. 
  \end{tabbing}
  \end{regime}

  This limit is such that we have a scaling of the fidelity as $F \gtrsim 1/\poly(n)$. It is also a natural scaling limit for noisy devices whose error rates gradually improve through better engineering. You might be worried about the fact that the depth needs to be quite low but it turns out that there is a solid quantum advantage even for depth-$\log n$ circuits.

  The precise definition of the weak-noise regime is motivated by  an observation that is crucial to assessing the noisy data from the experiment. 

  \subsection{Fidelity versus XEB: a phase transition}

  So let us turn to the second issue: Can we be convinced from the experimental data that finite-fidelity RCS has actually been solved?
  The quantity measured in the experiment is the XEB. Recall that the XEB averages the ideal probabilities corresponding to the sampled outcomes from experiments on random circuits. Thus, it correlates the experimental and ideal output distributions of those random circuits. You can think of it as a ``classical version'' of the fidelity: If the experimental distribution is correct, the XEB will essentially be 1, if it is uniformly random, the XEB is 0. 

  The experiments claimed that the XEB is a good proxy for the circuit-averaged fidelity $F$, and so we need to understand when this is true. Fortunately, in the past few years, alongside the improved experiments, we have developed a very good understanding \cite{dalzell_random_2024,barak_spoofing_2021,morvan_phase_2024,ware_sharp_2023}.

  It turns out that the quality of correspondence between XEB and average fidelity depends strongly on the noise in the experimental quantum state. In fact, there is a sharp phase transition: there is an architecture-dependent constant $c^*$ such that when the experimental noise is local with rate $\varepsilon < c^*/n$, then the XEB is a good and reliable proxy for the average fidelity for any system size $n$ and circuit depth $d$. This is exactly the weak-noise regime. Above that threshold, in the \emph{strong noise regime}, the XEB is an increasingly bad proxy for the fidelity \cite{morvan_phase_2024,ware_sharp_2023}. 

  Let me be more precise: In the weak-noise regime, when we consider the decay of the XEB as a function of circuit depth $d$, the \emph{rate of decay} is given by $\varepsilon n$, i.e., the XEB decays as $\exp(- \varepsilon n d )$. Meanwhile, in the strong-noise regime the rate of decay is constant, giving an XEB decay as $\exp(- \kappa d)$ for a constant $\kappa$. At the same time, the fidelity decays as $\exp(-\varepsilon n d)$ \emph{regardless of the noise regime}. Hence, in the weak-noise regime, the XEB is a good proxy of the fidelity, while in the strong noise regime, there is an exponentially increasing gap between the XEB (which remains large) and the fidelity (which continues to decay exponentially). 

\begin{figure}[h]
  \centering
  \includegraphics[width=.5\linewidth]{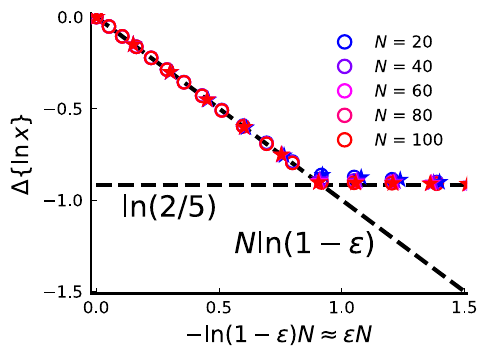}
  \caption{Depth-decay rate of the XEB as a function of normalized noise rate $\varepsilon N$, from \cite{ware_sharp_2023}. The $y$-axis label $\Delta( \ln \chi )$ is the decay rate of the XEB $\chi$, $N=n$ the number of qubits and $\varepsilon$ is the local noise rate. \label{fig:phase transition}}
  \end{figure}

  This is what \cref{fig:phase transition} shows. In Ref.~\cite{ware_sharp_2023}, we  computed it from an exact mapping of the behavior of the XEB to the dynamics of a statistical-mechanics model that can be evaluated efficiently. Using this mapping, we can also compute the noise threshold $c^*$ for whichever random circuit family and architecture you are interested in. 

  \paragraph{Where are the experiments?}

  We are now ready to take a look at the crux when assessing the noisy data: Can we trust the reported XEB values as an estimator of the fidelity? If so, do the experiments solve finite-fidelity RCS in the solidly hard regime where $\delta \geq 1/ \mathsf{poly}(n)$? 

  In their more recent paper \cite{morvan_phase_2024}, the Google team explicitly verified that the experiment is well below the phase transition, and it turns out that the first experiment was just at the boundary. The USTC experiments had comparable noise rates, and the Quantinuum experiment much better ones. Since fidelity decays as $\exp(-\varepsilon n d)$, but the reported XEB values stayed consistently around 0.1\% as $n$ was increased,  the experimental error rate $\varepsilon$ of the experiments improved even better than the $1/n$ scaling required for the weak-noise regime.
   Altogether, the experiments are in the weak-noise regime both in terms of absolute numbers and the required scaling.

  Of course, to derive the transition, we made some assumptions about the noise such as that the noise is local, and that it does not depend much on the circuit itself. In the advantage experiments, these assumptions about the noise are characterized and tested. This is done through a variety of means at increasing levels of complexity, including detailed characterization of the noise in individual gates, gates run in parallel, and eventually in a larger circuit. The importance of understanding the noise shows in the fact that a significant portion of the supplementary materials of the advantage experiments is dedicated to getting this right. This is experimental justification for using the XEB as a proxy for the fidelity. 

\begin{figure}[h]
  \centering
  \includegraphics[width=.7\linewidth]{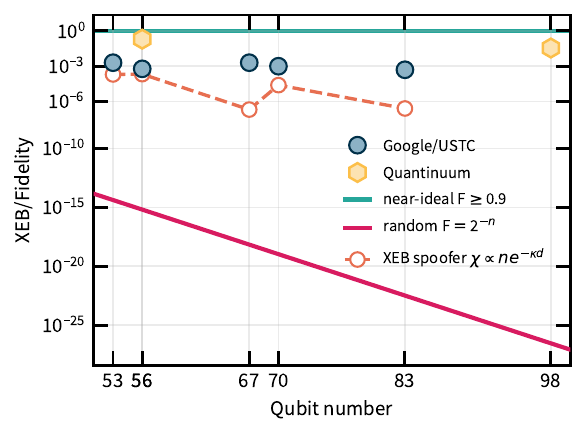}
  \caption{ \label{fig:xeb fidelity comparison}Fidelities of the experiments as a function of qubit number $n$ relative to the near-ideal score $F \ge 90$\% and the uniformly random score $F = 2^{-n}$ as well as the XEB score we expect from the spoofer by \textcite{gao_limitations_2024} for the Google/USTC circuits at depth $d$.}
  \end{figure}

  The data shows that the experiments solved finite-fidelity RCS for values of $\delta$ that stayed above the constant value of roughly 0.1\% even as the experiments grew. In \cref{fig:xeb fidelity comparison}, I compare the experimental fidelity values to the near-ideal scenario on the one hand, and the trivial $2^{-n}$ value on the other hand. Viewed at this scale, the values of $\delta$ for which the experiment solved finite-fidelity RCS are indeed vastly closer to the near-ideal value than the trivial baseline, which should boost our confidence that reproducing samples at a similar fidelity is extremely challenging.

  \paragraph{But we cannot compute the XEB!}
  You might be worried that the experiments did not actually compute the XEB in the advantage regime because to estimate it they would have needed to compute ideal probabilities---a task that is hard by definition of the advantage regime. Instead, they used several different ways to extrapolate the true XEB from XEB proxies (proxy of a proxy of the fidelity). Is this a valid way of getting an estimate of the true XEB? 

  It is.  Different extrapolations---from easy-to-simulate to hard-to-simulate, from small system to large system et cetera---all gave consistent answers for the experimental XEB value of the supremacy circuits. Think of this as having several lines that cross in the same point. For that crossing to be a coincidence, something conspiratorial must happen \emph{exactly} when you move to the supremacy circuits from different directions. That is why it is reasonable to trust the reported value of the XEB. 
  In fact, for the 2019 experiment \cite{arute_quantum_2019}, the XEB has in the meantime been computed using new simulation techniques, which has confirmed the estimated value within error bars \cite{liu_validating_2024}, lending further credence to the methodology.

  In contrast to the complexity of simulating noisy experiments, the complexity of computing the XEB does not reduce at lower fidelity of the samples. 
  This is because to verify an experiment from an empirical estimate of the XEB given the observed samples, we need to compute ideal probabilities. 
  Even to make verification and simulation runtimes match up, near-ideal fidelity is required.  
  This means that finite-fidelity experiments are often even easier to simulate than to verify using our best available method, the XEB.
  In all of the experiments, computing the XEB was out of reach and the physics methodology described in this section was necessary to give evidence that they achieved a certain value.

  \paragraph{That is exactly how experiments work!}
  Establishing that the experiments correctly solved finite-fidelity RCS and therefore show quantum advantage involved a lot of experimental characterization of the noise as well as theoretical work to understand the effects of noise on the quantity we care about---the fidelity between the experimental and ideal states. 
  In this respect (and maybe also in the scale of the discovery), the quantum advantage experiments are similar to the recent experiments reporting discovery of the Higgs boson and gravitational waves. While I do not claim that all the details and standards in the respective communities are the same, in both experiments, there is an unfathomable amount of data that could not be interpreted without preselection and post-processing of the data, theories, extrapolations and approximations that model the experiment and measurement apparatus, to take into account noise sources and the like. All of those enter the respective smoking-gun plots that show the discoveries. 

  If you believe in the validity of experimental physics methodology, you should therefore also believe in the type of evidence underlying the experimental claims of the quantum advantage demonstrations: that they sampled from the output distribution of a quantum state with the reported fidelities. 
  Put succinctly: 
  \begin{quote}
  \emph{If you believe in the Higgs boson and gravitational waves, you should probably also believe in the experimental demonstration of quantum advantage.} 
  \end{quote}

  \subsection{Exploits of the XEB phase transition and the complexity of X-RCS}

  You might be tempted to say: ``Well, but is all this really so important? Can we not just go back to considering the X-RCS task for the classical samples, and forget all about states and fidelity?''

  The phase transition shows why that would change the complexity of the problem: in the strong-noise regime, XEB can stay high even when fidelity is exponentially small. And indeed, this discrepancy can be exploited by so-called \emph{spoofers} for the XEB. These are efficient classical algorithms which succeed at a quantum advantage test even though they clearly do not achieve the intended advantage. These spoofers \cite{barak_spoofing_2021,gao_limitations_2024} can achieve high XEB scores that are on the same order of magnitude as those of the experiments and scaling like $\exp(-\kappa d)$ in the circuit depth~$d$ for some constant~$\kappa$. 
  Notice that these spoofers are very  different in type from the finite-fidelity sampling algorithms of \cite{pan_solving_2022,kalachev_classical_2021} in that they achieve a genuinely polynomial runtime scaling rather than just an improved exponential scaling. At the same time, their underlying simplification mechanism is the same, namely, to remove edges in the tensor-network diagram.

  Their basic idea is to introduce strong, judiciously chosen noise at specific circuit locations that has the effect of breaking up the simulation task into smaller, much easier components, but at the same time still gives a high XEB score. In doing so, they exploit the strong-noise regime in which the XEB is a \emph{really bad proxy} for the fidelity. This allows them to sample (in any product basis) from states with exponentially low fidelity while achieving a high XEB value. 

  The discovery of the phase transition and the associated spoofers highlights the importance of modeling when assessing---and even formulating---the advantage claim based on noisy data. 

  \paragraph{Scalable advantage for X-RCS?}
  Looking at the details more closely, the findings of the work on spoofing and the XEB phase transition provide a path to arguing that the noisy experiments demonstrate quantum advantage. 
  For this, let us look in more detail at the comparative behaviour of the experiments and the spoofers: 
  In the weak-noise and low-depth regime ($d=\ln n$), the experiments achieve an inverse polynomial scaling of the XEB as $n^{-\varepsilon n}$ with $\varepsilon n< c^*$, while the polynomial-time spoofers achieve a scaling as $n^{- \kappa}$. 
  Hence, both XEB signals decay inverse polynomially, but with \emph{a different power}. 
  While efficient spoofers are presumably limited by some fundamental bound determined solely by the capabilities of classical computers, in noisy quantum devices the noise rate can be (and has been) improved over the generations of experiments. 
  A reasonable conjecture is therefore that there is a classical lower bound $\kappa^*$ such that no efficient classical algorithm can achieve XEB score $\chi \gtrsim n^{-\kappa^*}$. 
  Conversely, log-depth noisy circuits with noise rate $\varepsilon < \kappa^*/n$ beat this bound and can therefore demonstrate quantum advantage for X-RCS. 

  In fact, this is the situation for the current noise rate and achievable spoofing decay rate.
  For the Google/USTC circuits, the spoofers of  \textcite{gao_limitations_2024} achieve a value of $\kappa$ somewhere between $0.4$ and $0.5$, while the experiments achieved a value of $\varepsilon n $ somewhere between $0.2$ and $0.4$.
  In \cref{fig:xeb fidelity comparison}, I also show the spoofer scores we expect for the Google/USTC experiments based on extrapolating the scores of the spoofer by \textcite{gao_limitations_2024} to the corresponding system sizes and depths. 
  Whether you believe that the current experiments demonstrate quantum advantage for X-RCS therefore depends on whether you believe that the current spoofers can be improved, or how much so. 

  More generally, we expect the computational threshold $\kappa^* \le c^*$ to be  below the XEB/fidelity transition point $c^*$ since a spoofer can choose bespoke, local noise whereas the phase transition is for uniform noise.  
  But $c^*$ is a property of a circuit ensemble that we can compute; $\kappa^*$ is a computational conjecture about the same ensemble. 
  While the experiments convincingly show that they lie below the XEB phase transition point, 
  it therefore remains  an open question whether they are also below the computational transition. \\ 

  So even as a cryptographer, if you want the evidence for quantum advantage to live up to a higher standard than that of experimental physics, you should probably believe that the noisy experiments solve a task (X-RCS) with scalable quantum advantage at some noise rate. Plausibly, even the noise rate of the latest experiments is sufficiently low.

  \paragraph{Complexity regions of RCS}
  Summarizing these findings, the complexity landscape of noisy RCS with XEB as a benchmark appears to decompose into three regions as a function of the noise. 
  The first region is the near-ideal regime at high fidelities, say, larger than $90\%$. In this regime, classical simulation of random circuits requires runtime scaling essentially as $2^{n}$ to the best of our knowledge for any of the RCS tasks outlined above. 
  The ion trap experiments plausibly approach this regime \cite{decross_computational_2025,ransford_98-qubit_2026}.

  The second regime is for local noise rates $\varepsilon < c^*/n$ below the phase transition $c^*$ of the XEB versus fidelity.
  Experiments in this regime can demonstrate quantum advantage for finite-fidelity RCS based on XEB because it is a good estimator of the fidelity. 
  The discoveries of Refs.~\cite{pan_solving_2022,kalachev_classical_2021} suggest that finite-fidelity RCS  at fidelity $\delta$ only reduces the complexity compared to ideal sampling by a factor of $\delta$. 
  In the first sets of experiments \cite{arute_quantum_2019,wu_strong_2021}, this could be exploited to make their classical simulation possible at the scale of the experiment. 
  In this regime, X-RCS may be spoofable by classical algorithms exploiting the high-noise regime. The experiments are convincingly in this regime and therefore demonstrate quantum advantage up to experimental physics standard!

  The third region lies below the ``computational'' phase transition, corresponding to the best possible $\kappa^*$ achievable by an efficient classical algorithm. 
  It is a reasonable conjecture that $\kappa^* > 0 $ and  therefore experiments with $\varepsilon < \kappa^*/n$ can demonstrate a quantum advantage for X-RCS in terms of a distinct polynomial scaling of the achieved XEB value. 
  The experiments may be in this regime, so they may even demonstrate quantum advantage up to a higher standard---no assumptions about states or noise characterization are needed, but experimental physics standards are still used to argue for the achieved \emph{value of the XEB}.

  \subsection{What are the counter-arguments?}

  \paragraph{The scaling hardliner}
  \begin{quote}
  \emph{``The weak-noise limit is not physical. The appropriate scaling limit is one in which the local noise rate of the device is constant while the system size grows, and in that case, there is a classical simulation algorithm for RCS \cite{bremner_achieving_2017,aharonov_polynomial-time_2023}.''}
  \end{quote}

  In theoretical computer science, scaling of time or the system size in the input size is considered very natural: We say an algorithm is efficient if its runtime and space usage only depend polynomially on the input size. But all scaling arguments are hypothetical concepts, and we only care about the scaling at relevant sizes. 

  In the end, every scaling limit is going to hit the wall of physical reality---be it the amount of energy or human lifetime that limits the time of an algorithm, or the physical resources that are required to build larger and larger computers. To keep the scaling limit going as we increase the size of our computations, we need innovation that makes the components smaller and less noisy. 

  At the scales relevant to RCS, the $1/n$ scaling of the noise is benign and even natural. Why? 
  Currently, the actual noise in quantum computers is not governed by the fundamental limit, but by engineering challenges. Realizing this limit therefore amounts to engineering improvements in the system size and noise rate that are achieved over time. At some point that scaling limit is also going to hit a fundamental barrier below which the noise cannot be improved. But we are surely not at that limit, yet. What is more, already now logical qubits are starting to work and achieve beyond-breakeven fidelities. So even if the engineering improvements should flatten out from here onwards, quantum error correction will keep the $1/n$ noise limit going and even accelerate it in the intermediate future. 

  \paragraph{The complexity maniac}
  \begin{quote}  
  \emph{``All the hard complexity-theoretic evidence for quantum advantage is in the near-ideal regime, but now you are claiming advantage for the low-fidelity version of that task.''}
  \end{quote}

  This is probably the strongest counter-argument in my opinion, and I gave my best response above. Let me just add that this is a question about computational complexity. In the end, all of complexity theory is based on belief. The only real evidence we have for the hardness of \emph{any task} is the \emph{absence} of an efficient algorithm, or the reduction to a paradigmatic, well-studied task for which there is no efficient algorithm. 

  Such evidence is hard to quantify, but at this point of sustained efforts aimed at developing classical simulations, I am convinced that there is no efficient algorithm for finite-fidelity RCS in the regime of the experiments.

  \paragraph{The enthusiastic skeptic}
  \begin{quote}
    
  \emph{``There is no verification test that just depends on the classical samples, is efficient and does not make any assumptions about the device. In particular, you cannot unconditionally verify fidelity just from the classical samples. Why should I believe the data?''}
  \end{quote}

  Indeed, the current advantage demonstrations are not device-independent. But the comparison you should have in mind are Bell tests. The first proper Bell tests of Aspect and others in the 1980s were not free of loopholes. They still allowed for contrived explanations of the data that did not violate local realism. Still, I can hardly believe that anyone would argue that Bell inequalities were not violated already back then. 

  As the years passed, these remaining loopholes were closed. To be a skeptic of the data, people needed to resort to more and more adversarial scenarios that could explain the data. We are working to achieve the same with quantum advantage demonstrations: devise better schemes and better tests that require fewer and fewer assumptions or knowledge about the specifics of the device. 

  \paragraph{The task purist}
  \begin{quote}
  \emph{``When you chose the gates and architecture of the circuit dependent on your device, you tailored the task too much to the device and that is unfair. Not even the different RCS experiments solve exactly the same task.''}
  \end{quote}

  The specific computations solved are indeed tailored to the hardware itself and the individual gates are only specified after the calibration and characterization of the chip. 
  In this sense the experiments are not hardware-independent yet. 
  However, the characterization is a separate experiment, and once it is done, the gate set is fixed and can be used to compile any unitary. 
  Thus, I argue that the experiments still solve fine computational tasks, since the circuit description is supplied at the beginning of each run. 
  Moving away from such hardware-tailored task specifications is an important next step, however. 

  \paragraph{The task historian}
  \begin{quote}
  \emph{``When you define finite-fidelity RCS in terms of the ability to perform any Pauli measurement, this is a post-hoc adjustment to the task that was never claimed in the original papers.''}
  \end{quote}

  This point is well taken. Indeed, for example, the 2019 paper states explicitly that the task is to ``sample the output of a pseudo-random quantum circuit'' with the verification  goal to ``achieve a high enough [XEB] (...) such that the classical computing cost is prohibitively large.'' \cite[p.~506]{arute_quantum_2019}. 
  While this standard has been held up for the experiment, the claim that XEB alone \emph{scalably} witnesses advantage or correct samples is---in my opinion---not finally settled because of the new spoofing algorithms.
  At the same time, the experiments performed significantly more benchmarking of the noise and theory of XEB relating to fidelity to justify that in fact they prepared a quantum state with high fidelity. 
  This is very explicit in the Quantinuum papers \cite{decross_computational_2025,ransford_98-qubit_2026}. 

  In the end, we should care about the question what task with a quantum advantage noisy experiments can be said to solve.
  What we have learned since the first experiments were performed should enter our assessment here. 
  I argued that finite-fidelity RCS is the notion that most meaningfully interprets the experiments, while at the same time allowing for reasonable classical simulation attempts. 
  It gives a coherent story that fits all the pieces in the supremacy claims and theoretical work around them. 
  And quantum advantage for X-RCS is still not ruled out.   
  \\

  Finally, let me bring up a skeptical argument coming from the opposite perspective. 

  \paragraph{The contented pragmatist} 
  \begin{quote}
    \emph{``You are making life too difficult for yourself with such a complicated argument. Quantum advantage just means that classical algorithms take exponential time. By this standard, the 2019 experiment already settled the question once and for all.''}
  \end{quote}
  To dissect this position, we first need to ask what ``exponential time'' refers to. 
  Read as a scaling claim, the 2019 experiment indeed settled it, and no algorithms with a subexponential runtime have been found in the meantime. 
  But any single experiment only ever solves an instance of a problem, and a scaling claim is always a property of a family of increasing problem sizes. 
  So the scaling interpretation begs the question of where to draw the line for quantum advantage---by a mere scaling standard, already a 5-qubit experiment would demonstrate quantum advantage. 

  This is why my definition of quantum advantage included the criterion that the solved task is ``hard in practice (at the time)''.\footnote{The term ``quantum supremacy'' was introduced to separate experiments satisfying this condition from ``mere'' quantum advantage. But the distinction has become blurry over the years.  } 
  This requirement is inherently vague as what we count as being possible at the time depends on how much time we give people to actually try and simulate an experiment. 
  While the 2019 experiment was simulated after two years \cite{pan_solving_2022}, today the samples can be matched within minutes \cite{zhao_leapfrogging_2025}. 
  So both criteria are needed: The required size of an experiment is fixed by the in-practice claim; the scaling claim makes it survive progress in algorithm design. 

  \subsection{Quantum advantage with loopholes}

  At this point, I have given you my arguments for why I am convinced quantum advantage has been achieved. 
  In particular, I have argued for the following resolution to the questions I raised at the beginning of this section: 
  Quantum advantage experiments using noisy random circuit sampling solve finite-fidelity RCS which is a ``CS-style'' task. 
  I have argued that we have strong evidence for asymptotic and in-practice hardness of that task and that the evidence that the experiments actually solve that task lives up to the standards of experimental physics. 
  This leaves open the possibility of yet stronger evidence for hardness, more scalable experiments, and verification up to a higher standard that is used, say, in cryptography. 
  The analogy to keep in mind is the  Bell inequality violations of the 1980s with loopholes that are yet to be closed. 

  Where does this leave us? 
  The stated theoretical motivations for performing a quantum advantage experiment are to test whether quantum mechanics is still valid in the regime of high complexity, and to test whether some correlated noise will destroy any advantage. 
  If you do not believe in nature acting adversarially, then indeed the quantum advantage experiments have passed these tests---all our quantum mechanical models of the experiments, including the noise, predict the observed experimental outcomes. 
  We have not seen any deviations originating in physics appear in the regime of high complexity. 
  Under this threat model, the experiments demonstrate that quantum advantage is physically possible, and therefore fulfill their promise in the same way that the first Bell-inequality violations did. 

  Of course, the existence of loopholes always leaves room for nature acting adversarially.
  In the following section, I address why closing these loopholes is a worthwhile next step in quantum advantage for the era of 100 logical qubits. The most important---and theoretically interesting---one is to enable efficient verification of quantum advantage using less or even no specific knowledge about the device that was used, but just the measurement outcomes.

  \section{What is next in quantum advantage?}
  \label{sec:part3}

  We are now at an exciting point in our process of developing quantum computers and understanding their computational power: It has been demonstrated that quantum computers \emph{can} outperform classical ones (if you accept my argument from \cref{sec:part2}). And it has been demonstrated that quantum fault-tolerance is possible for at least a few logical qubits. Together, these form the elementary building blocks of useful quantum computing. 

  And yet: the devices we have seen so far are still nowhere near being useful for any advantageous application in, say, condensed-matter physics or quantum chemistry, which is where the promise of quantum computers lies. 
  And while the resource estimates for running cryptographically relevant algorithms have reduced dramatically in the past years \cite{gidney_how_2025,babbush_securing_2026,cain_shors_2026}, the required processor sizes are still in the thousands of logical qubits and billions of logical gates. 

  \begin{quote}
    \emph{So what is next in quantum advantage? }
  \end{quote}

  What are milestones we should work towards both in theory and experiment for the next generation of quantum processors, en route to eventually running large-scale cryptographically relevant quantum algorithms? 
  Today's physical, noisy processors operate on low hundreds of qubits and are capable of executing thousands of gates before noise drowns the signal. 
  At the same time, first long-lived single logical qubits encoded in the surface code have been built \cite{computing_quantum_2026}. 
  A computationally meaningful next generation of quantum processors to have in mind is the regime of 100 logical qubits:

  \begin{regime}[The 100 logical qubits regime]
  In this regime, there are 100 logical qubits with error rates on the order of $10^{-6}$, that is, 100 qubits on which we can run up to 1 million gates.
  \end{regime}

  Building devices operating in this regime will require 1\,000 to 10\,000 physical qubits and is therefore well beyond the proof-of-principle quantum advantage and fault-tolerance experiments that have been done.
  This is a regime that can presumably be reached with current technology using single processing units in ion traps  \cite{tripier_fault-tolerant_2026} or neutral atoms \cite{cain_shors_2026,menssen_strategic_2026}. 
  At the same time, it is  still orders of magnitude away in both time and space from any of the first cryptographic applications, for example in breaking elliptic-curve cryptography \cite{babbush_securing_2026}. Large-scale devices with several thousand logical qubits that allow us to tackle this regime may well require interconnected processing units. 
  At the lower end of the 100-logical-qubit regime, we might have access to a sub-universal gate set, while at the higher end, we will be capable of running universal computations~\cite{tripier_fault-tolerant_2026}.  
  And yet, there is not a clear picture for what we can and should do with such devices. 
  In the following, I argue, however, that it is a regime in which we can hope to do qualitatively new things.

  \paragraph{The next milestone: classically verifiable quantum advantage}

  In this section, I want to argue that a key milestone we should aim for in the 100 logical qubit regime is \emph{classically verifiable quantum advantage}. Achieving this will not only require the jump in quantum device capabilities but also finding advantage schemes that allow for classical verification using these limited resources. 

  \begin{quote}
    Why is this an interesting and feasible goal and what is it anyway? 
  \end{quote}

  \begin{figure}[t]
  \centering
  \includegraphics{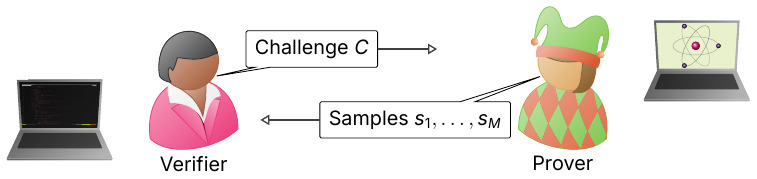}
  \caption{The theoretical computer scientist's cartoon of verifying a quantum computer from classical samples: a classical verifier (the user) challenges a quantum prover (the quantum computer) with a description of a circuit, and the prover returns measurement results.\label{fig:verifier prover} }
  \end{figure}
  \begin{figure}[t]
  \centering
  \includegraphics[width=.8\linewidth]{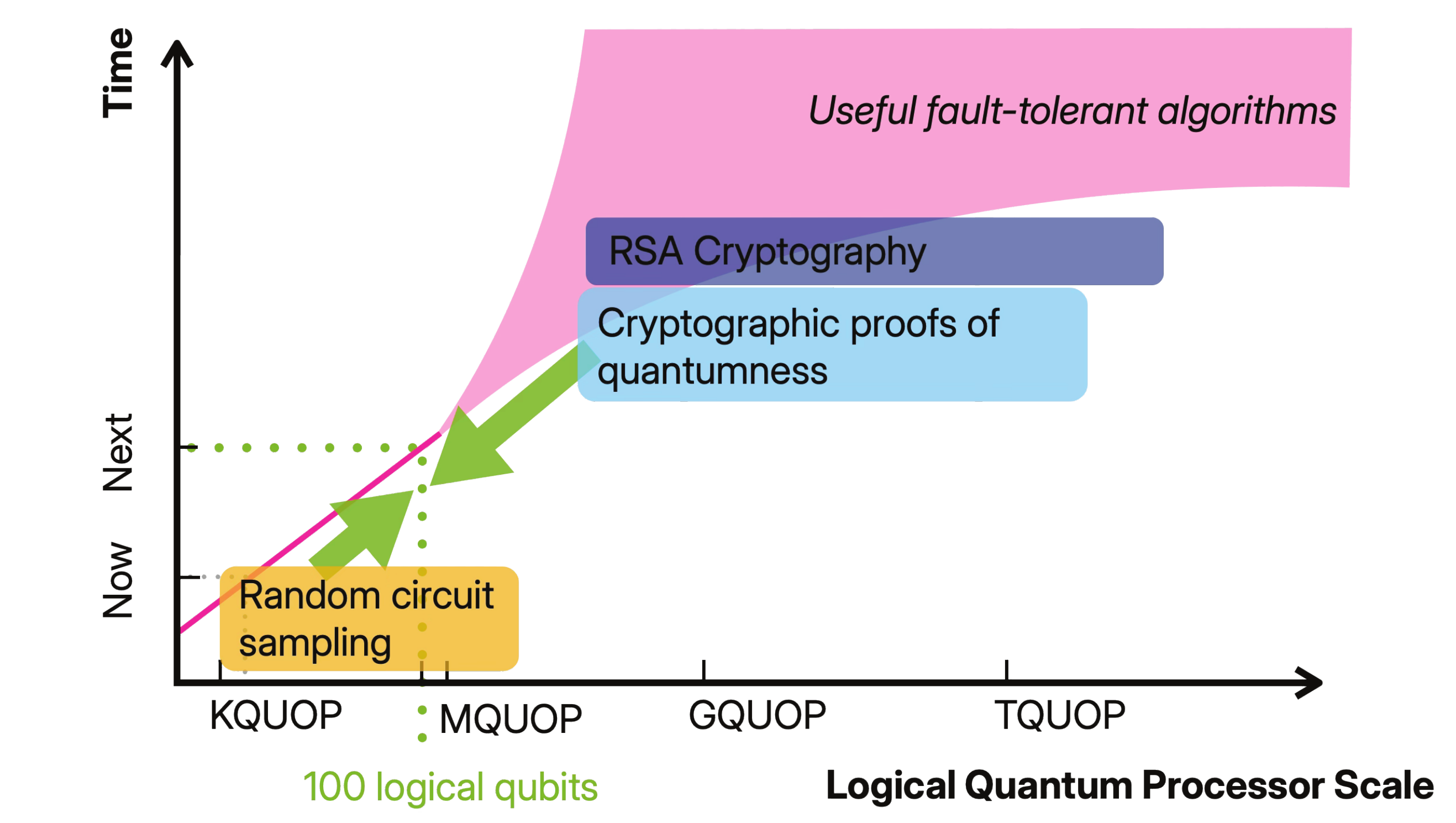}
  \caption{To achieve verifiable advantage in the 100-logical-qubit regime we need to close the gap between random circuit sampling and proofs of quantumness.\label{fig:algorithm landscape}}
  \end{figure}

  To my mind, the biggest weakness of the RCS experiments is the way they are verified. Currently, verification uses XEB which can be classically spoofed, and only actually measured in the simulatable regime. Really, in a quantum advantage experiment I would want there to be an efficient procedure that will without any reasonable doubt convince us that a computation \emph{must have} been performed by a quantum computer when we run it. In \emph{classically verifiable quantum advantage}, a (classical) verifier would devise challenge circuits which they would then send to a quantum server, as sketched in \cref{fig:verifier prover}. These would be designed in such a way that once the server returns \emph{classical} samples from those circuits, the verifier can convince herself that the server must have run a quantum computation.
  This could be done in a way that uses some secret the verifier holds and used to design the circuit (called private verification), 
  or even better in a way that \emph{anyone} given access to the input and outcome of the computation would be able to verify it (called public verification).

  This is the jump from a physics-type experiment (the sense in which advantage has been achieved) to a secure protocol that can be used in settings where I do not want to trust the server and the data it provides me with. Such security may also allow a first application of quantum computers: to generate random numbers whose genuine randomness can be certified---a task that is impossible classically. 

  Here is the problem: On the one hand, we do know of schemes that allow us to classically verify that a computer is quantum and generate random numbers, so-called \emph{cryptographic proofs of quantumness} \cite{brakerski_cryptographic_2018}. A proof of quantumness is a highly reliable scheme in that its security relies on well-established cryptography. Their big drawback is that they require a large number of qubits and operations, comparable to the resources required for factoring \cite{kahanamoku-meyer_classically_2022,kahanamoku-meyer_jacobi_2024,babbush_securing_2026}. On the other hand, the computations that we \emph{can} run in the advantage regime---essentially, random circuits---are very resource-efficient but not verifiable.  

  The 100-logical-qubit regime lies right in the middle, and it seems more than plausible that classically verifiable advantage is possible in this regime. The theory challenge ahead of us is to find it: a quantum advantage scheme that is very resource-efficient like RCS and also classically verifiable like proofs of quantumness. This idea is illustrated in \cref{fig:algorithm landscape}.

  With this in mind, let me spell out some concrete milestones that we can hope to achieve using 100 logical qubits on the road to classically verifiable quantum advantage. 

  \subsection{Demonstrate fault-tolerant quantum advantage}

  Before we talk about verifiable advantage, the first experiment I would like to see is one that combines the two big achievements of the past years, and shows that quantum advantage and fault-tolerance can be achieved simultaneously. Such an experiment would be similar in type to the RCS experiments, but run on encoded qubits with gate sets that match that encoding. During the computation, noise would be suppressed by correcting for errors using the code. In doing so, we could reach the near-perfect regime of RCS as opposed to the finite-fidelity regime that current RCS experiments operate in (as I discussed in detail in \cref{sec:part2}).   

\begin{figure}[h]
  \centering
  \includegraphics[width=.7\linewidth]{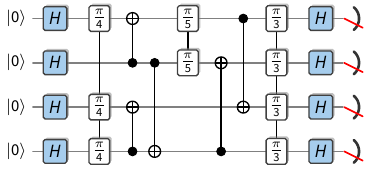}
  \caption{An IQP circuit starts from the all-$\ket 0$ state by applying a Hadamard transform, followed by diagonal and CNOT gates (in this case $Z(\pi/4,1011)$, some CNOT gates, $Z(\pi/5,1100)$, some CNOT gates, $Z(\pi/3,1111)$) and ends in a measurement in the Hadamard basis.
  \label{fig:iqp circuit}}
  \end{figure}

  Random circuits with a quantum advantage that are particularly easy to implement fault-tolerantly are so-called \emph{IQP circuits}. In those circuits, the gates are controlled-NOT gates and diagonal gates, so rotations $Z(\theta,z)$, which just add a phase to a basis state $\ket x$ as $Z(\theta, z) \ket x = \exp(i \theta \, z \cdot x) \ket x$. The only ``quantumness'' comes from the fact that each input qubit is in the superposition state $\ket + =  (\ket 0 + \ket 1)/\sqrt 2$, and that all qubits are measured in the $X$ basis. \cref{fig:iqp circuit} shows an example of an IQP circuit. 

  It turns out that IQP circuits are already really well understood since one of the first proposals for quantum advantage was based on IQP circuits \cite{shepherd_temporally_2009,bremner_classical_2010}, 
  and for a lot of the results in random circuits, we have precursors for IQP circuits, in particular, their ideal and noisy complexity \cite{bremner_average-case_2016,bremner_achieving_2017}. This is because their near-classical structure makes them relatively easy to study. Most importantly, their outcome probabilities are simple (but exponentially large) sums over phases $e^{i \theta }$ that can just be read off from which gates are applied in the circuit and we can use well-established classical techniques like Boolean analysis and coding theory to understand those. 

  IQP gates are natural for fault-tolerance because there are codes in which all the operations involved can be implemented \emph{transversally}. This means that they only require parallel physical single- or two-qubit gates to implement a logical gate. This is in stark contrast to universal circuits which require complicated and resource-intensive fault-tolerant protocols. Running computations with IQP circuits would also be a step towards running `real' algorithms in that they can involve structured components such as cascades of CNOT gates and the like. These show up all over fault-tolerant constructions of algorithmic primitives such as arithmetic or phase estimation circuits.

  Our concrete proposal for an IQP-based fault-tolerant quantum advantage experiment in reconfigurable-atom arrays is based on interleaving diagonal gates and CNOT gates to achieve super-fast scrambling \cite{hangleiter_fault-tolerant_2025}. A medium-size version of this protocol was implemented by the Harvard group \cite{bluvstein_logical_2024} but with only a bit more effort, it could be performed in the advantage regime. 
  Reaching this regime will require larger qubit numbers than for random circuits since more efficient simulation algorithms are available. In the worst case, this might require doubling the qubit number \cite{dalzell_how_2020}, which means we are exactly in the regime of 100 logical qubits. 

  In those proposals, verification will still suffer from the same problems as standard RCS experiments, so the task ahead of us is to fix that! 

  \subsection{Closing the verification loophole}

  The milestone I proposed for the 100-logical-qubit regime is to find schemes that lie in between RCS and proofs of quantumness in terms of their resource requirements but at the same time allow for more efficient and more convincing verification than RCS. Naturally, there are two ways to approach this space---we can make quantum advantage schemes more verifiable, and we can make proofs of quantumness more resource-efficient. 

  First, let us focus on the former approach and set a more moderate goal than secure classical verification of data from an untrusted server. Are there variants of RCS that allow us to efficiently verify that finite-fidelity RCS has been achieved if we trust the experimenter and the data they hand us? 

  \subsubsection{Efficient quantum verification using random circuits with symmetries}

  Indeed, there are. I think of the schemes that achieve this as \emph{random circuits with symmetries}. A symmetry is an operator $S$ such that the outcome state of the computation $\ket C$ (or some intermediate state) is invariant under the symmetry, so $S \ket C =\ket C$. The idea is then to find circuits $C$ that exhibit a quantum advantage and at the same time have symmetries that can be easily measured, say, using only single-qubit measurements or a single gate layer. Then, we can use these measurements to check whether or not the pre-measurement state respects the symmetries. This is a test for whether the quantum computer prepared the correct state, because typical errors or deviations from the true state would violate the symmetry (unless they were adversarially engineered).
  In other words, the symmetries give rise to an \emph{error detecting code}.
  We can even devise circuits for which the expectation values of the symmetries determine the full state fidelity, and we might add mid-circuit measurements to aid in this.

  In random circuits with symmetries, we can thus use small, well-characterized measurements whose outcomes we trust to probe whether a large quantum circuit has been run correctly. 
  This is exactly the physicists' scenario in which experimental methodology can be used. 
  I call it the \emph{trusted experimenter scenario}.

  \begin{scenario}[Trusted experimenter]
  In this scenario, we receive data from an actual, potentially noisy experiment in which we trust that certain measurements were actually and correctly performed.
  \end{scenario}

  \begin{figure}[h]
  \centering
  \includegraphics[width=.5\linewidth]{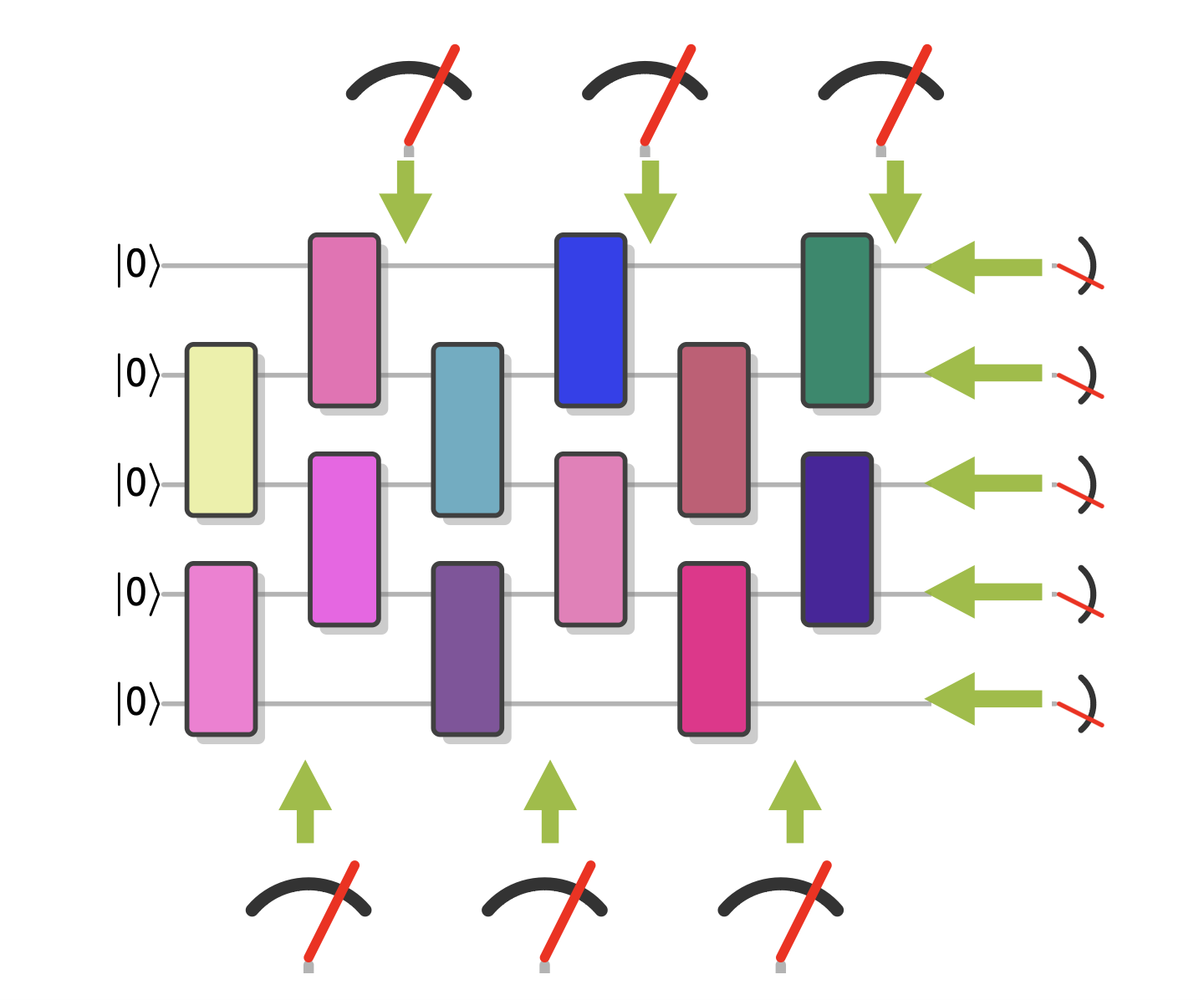}
  \caption{Random circuits with symmetries  introduce measurements in the circuit that check for errors but do not disturb the ideal computation.}
  \end{figure}

  Here are some examples of random circuits with symmetries, which allow for efficient verification of quantum advantage in the trusted experimenter scenario.

  \paragraph{Graph states} My first examples are locally rotated graph states \cite{ringbauer_verifiable_2025}. These are states that are prepared by CZ gates acting according to the edges of a graph on an initial all-$\ket +$ state, and a layer of single-qubit $Z$-rotations is performed before a measurement in the $X$ basis. (This is also an IQP circuit.) The symmetries of this circuit are locally rotated Pauli operators, and can therefore be measured using only single-qubit rotations and measurements. What is more, these symmetries fully determine the graph state. Determining the fidelity then just amounts to averaging the expectation values of the symmetries. In this example, measuring the outcome state to obtain hard-to-reproduce samples and measuring the symmetries are done in two different (single-qubit) bases.

  To achieve quantum advantage, graph states on a highly connected graph on one to a few hundred qubits should be sufficient. These could be  generated on 100 logical qubits using intermediate measurement and reset of qubits.

  \paragraph{Bell sampling} The drawback of this approach is that we need to make two different measurements for verification and sampling. But it would be much neater if we could just verify the correctness of a set of classically hard samples by \emph{only using those samples}. For an example where this is possible, consider two copies of the output state $\ket C$ of a random circuit, so $\ket C \otimes \ket C$. This state is invariant under a swap of the two copies, and in fact the expectation value of the SWAP operator $\mathbb S$ on two copies of a noisy state preparation $\rho_C$ of $\ket C $ determines the purity of the state, so $P
   = \mathrm{tr}[ (\rho_C \otimes \rho_C) \mathbb S]$. It turns out that sampling the outcomes of measurements of all pairs of qubits in the state $\ket C \otimes \ket C$ in the pairwise basis of the four Bell states $\ket{\sigma} = (\sigma \otimes \id) (\ket{00} + \ket{11})/\sqrt 2$, where $\sigma$ are the four Pauli matrices $\id, X, Y, Z$, is hard to simulate classically 
  \cite{hangleiter_bell_2024}. You may also observe that the SWAP operator is diagonal in the Bell basis, so its expectation value can be extracted from the Bell-basis measurements---our hard to simulate samples. To do this, we just average sign assignments to the sampled bitstrings according to their parity. 

  If the circuit is random, then under the same assumptions as those used in XEB for random circuits, the root purity is a good estimator of the fidelity of each copy, so $\sqrt P \approx F(C)$. So this is an example where \emph{efficient} verification is possible directly from hard-to-simulate classical samples under the same assumptions as those used to argue that XEB equals fidelity. 

  To achieve quantum advantage using Bell sampling, intuitively, we must roughly double the number of qubits compared to universal random circuit sampling, since we need two copies of a classically hard circuit. 
  This could be achieved with just over 100 logical qubits. 

  \paragraph{Fault-tolerant circuits} Finally, suppose that we run a fault-tolerant quantum advantage experiment. Then, there is a natural set of symmetries of the state at any point in the circuit, namely, the stabilizers of the code we use. In a fault-tolerant experiment we repeatedly measure those stabilizers mid-circuit, so why not use that data to assess the quality of the logical state? Indeed, it turns out that the logical fidelity can be estimated efficiently from stabilizer expectation values even in situations in which the logical circuit has a quantum advantage~\cite{xiao_situ_2026}. 
  This setting is particularly appealing because it can combine closing the scalability loophole and the verifiability loophole in the trusted experimenter scenario. 

  IQP circuit sampling is likely to require at most double the qubit count compared to universal random circuits \cite{dalzell_how_2020}, so to achieve quantum advantage with fault-tolerant IQP circuits\cite{hangleiter_fault-tolerant_2025} roughly 100 logical qubits will be sufficient. 

  In fact, very recently, \textcite{martiel_sampling_2026} developed a near-term idea along these lines based on deliberately adding symmetries and associated mid-circuit measurements that correspond to stabilizer measurements in a certain space-time code.
  Using this scheme, they were able to perform large-scale computations with high fidelity on noisy physical processors.
  The in-practice advantage of their experiment remains outstanding: The concrete circuits were simulated quickly after \cite{manabe_classical_2026} to which the authors responded by slightly modifying the circuits to circumvent this simulation  algorithm.
  Either way, their experiment shows: identifying codesigned circuit ensembles with symmetries that enable fidelity estimation is indeed feasible and rewarding in the near term.\\ 

  In all of these examples of random circuits with symmetries, we can show classical hardness in the trusted-experimenter scenario and the verification is public. However, coming up with classical samples that pass the verification tests is very easy, so the trust assumption on the  experimenter is crucial for this to work. (Note, however, that it may be possible to add tests to Bell sampling that make spoofing difficult \cite{hangleiter_validated_2023}.) At the same time, these proposals are very resource-efficient in that they only increase the quantum cost compared to a pure random-circuit experiment with the same in-practice classical complexity by a relatively small amount. 
  What is more, the required circuits have more structure than random circuits in that they typically require gates that are natural in fault-tolerant implementations of quantum algorithms. 

  Performing random circuit sampling with symmetries is therefore a natural next step en route to both classically verifiable advantage that closes the \emph{no-efficient verification} loophole, and towards implementing actual algorithms.

  What if we do not want to afford that level of trust in the person who runs the quantum circuit, however? 

  \subsubsection{Classical verification using random circuits with planted secrets}

  As cryptographers, we might not want to trust the experimenter, and we are in the \emph{untrusted quantum server scenario}.

  \begin{scenario}[Untrusted quantum server]
  In this scenario, we delegate a quantum computation to an untrusted (presumably remote) quantum server---think of using a cloud server to run your computation. We can communicate with this server using classical information.
  \end{scenario}

  In the untrusted server scenario, we can hope to use ideas from proofs of quantumness such as the use of classical cryptography to design families of quantum circuits in which some secret structure is planted that remains private knowledge of the verifier. This secret structure should give the verifier a way to check whether a set of samples passes a certain verification test. 
  At the same time it should not be detectable, or at least not be identifiable from the circuit description alone. 

  A simple example of a secret structure for private verification is a large peak in an otherwise flat output distribution of a random-looking quantum circuit. 
  The idea is that the verifier would pick a (random) string $x$ and design a circuit such that the probability of sampling $x$ is large. If the peak is hidden well, finding it just from the circuit description might require searching through all of the $2^n$ outcome bit strings and even just determining one of the outcome probabilities is exponentially difficult. A classical spoofer trying to fake the samples from a quantum computer would then be caught immediately: the list of samples they hand the verifier will not even contain $x$ with overwhelming probability, since there are exponentially many possible choices of $x$. 

  Unfortunately, planting such secrets seems to be very difficult using universal circuits, since the output distributions are so unstructured. This is why we have not yet found good candidates of circuits with peaks, but some attempts have been made \cite{aaronson_verifiable_2024,deshpande_peaked_2025,gharibyan_heuristic_2025}.
  We do have a promising candidate, though---IQP circuits. The fact that the output distributions of IQP circuits are quite simple could very well help us design sampling schemes with hidden secrets. Indeed, the idea of hiding peaks has been pioneered by Shepherd and Bremner \cite{shepherd_temporally_2009} who found a way to design classically hard IQP circuits with a large hidden Fourier coefficient. The presence of this large Fourier coefficient can easily be checked from just a handful of classical samples, and random IQP circuits do not have any large Fourier coefficients. Unfortunately, for that construction and a variation thereof \cite{bremner_instantaneous_2025}, it turned out that the large coefficient can be detected quite easily from the circuit description \cite{kahanamoku-meyer_forging_2023,gross_secret-extraction_2025}. 

  Choosing a different route than peaked circuits, we recently discovered that one can hide structure in Boolean polynomials that makes the simulation of corresponding IQP circuits easier than for random polynomials while keeping their low-moment statistical properties unchanged \cite{blaser_verifiable_2026}. 
  This can be leveraged for verifiable quantum advantage using an XEB test on IQP circuits: under two fine-grained complexity assumptions, one about a natural generalization of the planted clique problem \cite{jerrum_large_1992,hirahara_planted_2024}, the other about passing XEB for random IQP circuits, an adversary would have no other way of passing the test than simulating the ideal circuit at the average-case cost, whereas a verifier can estimate the XEB more efficiently using the secret structure. 
  However, while there is a large gap between verification and simulation times using classical computers, verification remains exponential. 
  In another IQP circuit sampling scheme, we obtain  efficient verification  based on hiding an efficiently decodable error-correcting code in a larger subspace \cite{deshpande_efficiently_2026}. 
  However, the security assumptions required for the scheme to have a quantum advantage are tailored to the scheme and remain not well tested yet. 
  To this day, it thus remains an open question whether secrets can be securely planted in (IQP) circuit families in a way that allows for \emph{efficient} classical verification.

  \paragraph{Towards applications: certifiable random number generation}

  Beyond verified quantum advantage, random circuit sampling schemes with hidden secrets may be usable to generate classically certifiable random numbers: You sample from the output distribution of a random circuit with a planted secret, and verify that the samples come from the correct distribution using the secret. If the distribution has sufficiently high entropy, truly random numbers can be extracted from them. The same can be done for RCS, except that some acrobatics are needed to get around the problem that verification is just as costly as simulation \cite{aaronson_certified_2023,liu_certified_2025}. 
  A large gap between verification and simulation times might permit such certified random number generation \cite{blaser_verifiable_2026}. 

  The goal here is firstly a theoretical one: Identify a planted-secret RCS scheme that has a large verification-simulation gap. But then, of course, it is an experimental one: actually perform such an experiment to classically verify quantum advantage. 

  Should an IQP-based scheme of circuits with secrets exist, 100 logical qubits is the regime where it should give a relevant advantage. 

  \subsection{Three milestones}

  Altogether, I have proposed three milestones for the 100 logical qubit regime. 
  \begin{goal}[Targets for 100 logical qubits] \quad \\ \vspace{-3ex}
  \begin{enumerate}
  \item Perform fault-tolerant quantum advantage using random IQP circuits. This will allow an improvement of the fidelity towards performing near-perfect RCS and thus close the scalability worries of noisy quantum advantage I discussed in \cref{sec:part2}. 

  \item Perform RCS with symmetries. This will allow for efficient verification of quantum advantage in the trusted experimenter scenario and thus make a first step toward closing the verification loophole. 

  \item Find and perform RCS schemes with planted secrets. This will allow us to verify quantum advantage in the remote untrusted server scenario and presumably give a first useful application of quantum computers to generate classically certified random numbers. 
  \end{enumerate}
  \end{goal}

  All of these experiments are natural steps towards performing more broadly useful quantum algorithms in that they use more structured circuits than just random universal circuits and can be used to benchmark the performance of the quantum devices in an advantage regime. Moreover, all of them close some loophole of the previous quantum advantage demonstrations, just like follow-up experiments to the first Bell tests have closed the loopholes one by one. 

  I have argued that IQP circuits will play an important role in achieving those milestones since they are a natural circuit family in fault-tolerant constructions and promising candidates for random circuit constructions with planted secrets. Developing a better understanding of the properties of the output distributions of IQP circuits will help us achieve the theory challenges ahead.  

  Experimentally, the 100 logical qubit regime is exactly the regime to target with those circuits since while IQP circuits are somewhat easier to simulate than universal random circuits, 100 qubits is well in the classically intractable regime. 

  \subsection{Outlook }

  Let me close  by touching on some directions that warrant more extensive discussion than this perspective permits. 

  First, there is the OTOC experiment by the Google team \cite{abanin_observation_2025} which has spawned quite a debate. This experiment claims to achieve quantum advantage for an arguably more natural task than sampling, namely, computing expectation values. Computing expectation values is at the heart of quantum-chemistry and condensed-matter applications of quantum computers. And it has the convenient property that it is what the Google team called ``quantum-verifiable'' in the following sense: Suppose we perform an experiment to measure a classically hard expectation value on a noisy device now, and suppose this expectation value actually carries some signal, so it is significantly far away from zero. Once we have a trustworthy quantum computer in the future, or even two distinct quantum computers able to solve exactly the same task, we will be able to check that the outcome of this experiment was correct and hence quantum advantage was achieved. 
  In contrast, all notions of verifiability I discussed in this perspective only require classical computers. 

  There is also an interesting theory challenge that relates to the noise-scaling arguments I discussed in detail in \cref{sec:part2}: The challenge is to understand whether quantum advantage can be achieved in the presence of a constant amount of local noise. What do we know about this? On the one hand, log-depth random circuits with constant local noise are easy to simulate classically \cite{bremner_achieving_2017,aharonov_polynomial-time_2023}, and we have good numerical evidence that random circuits at very low depths are easy to simulate classically even without noise \cite{napp_efficient_2022}. So is there a depth regime in between the very low depth and the log-depth regime in which quantum advantage persists under constant local noise? Is this maybe even true in a noise regime that does not permit fault-tolerance (see for example this talk \cite{aharonov_quantum_2025})? In the regime in which fault-tolerance is possible, it turns out that one can construct simple fault-tolerance schemes that do not require any quantum feedback \cite{nguyen_polylog_2025}, so there are distributions that are hard to simulate classically even in the presence of constant local noise. 

  \section*{Conclusion}

  In this perspective, my first goal was to convince you that quantum advantage has been achieved. There are some open loopholes, but if you are happy with physics-level experimental evidence, then I hope you are convinced that the RCS experiments of the past years have demonstrated quantum advantage.

  As the devices are getting better at a rapid pace, there is a clear goal that I hope will be achieved in the 100-logical-qubit regime: demonstrate fault-tolerant and verifiable advantage (for the experimentalists) and develop  the schemes to do that (for the theorists). Those experiments would close loopholes of the current RCS experiments. And they would work as a stepping stone towards algorithms with cryptographic applications in the advantage regime.

  \section*{AI statement}
  In this article, I have expressed my perspective on the status and future of quantum advantage. 
  The voice, thoughts, and judgement you read are my own and no wording has been touched by a large language model. 
  ChatGPT 5.2, 5.6 and 6, Claude Sonnet 4.6, Opus 5 and Fable 5 have been helpful tools for giving me feedback on the structure, framing and technical inconsistencies of the manuscript in the late stages of the writing process. 
  In particular, they made me aware of the (admittedly very simple) counterexample to my original finite-fidelity task statement \cite{hangleiter_blog_2026} that necessitated the refinements I present.

  \section*{Acknowledgements} 
  This perspective was published originally as a mini-series in three parts on the Caltech \emph{Quantum Frontiers} blog~\cite{hangleiter_blog_2026}, and I owe a big ``Thank you!'' to Spiros Michalakis, John Preskill, and Frederik Hahn who have patiently read and greatly helped me improve it. 
  I am also immensely grateful to Abhinav Deshpande, Yunchao Liu, and Michael Foss-Feig for discussions from which my framing of the various RCS tasks and their advantage has emerged. 
  An anonymous referee and Katiuscia Cassemiro have provided engaging feedback that has helped immensely to sharpen this perspective.  
  Finally, I am grateful to my collaborators and friends over the years for the many discussions without which this perspective would not be thinkable, most importantly, Michael Gullans,  Abhinav Deshpande, Bill Fefferman, and Soumik Ghosh. 
  I was supported by a Simons postdoctoral fellowship through DOE QSA and NSF QLCI Grant No.\ 2016245, and by the Swiss National Science Foundation through Ambizione Grant No.\ 223764.

  \printbibliography
  \end{document}